\documentclass[conference]{IEEEtran}
\usepackage{cite}
\usepackage{amsmath,amssymb,amsfonts}
\usepackage{algorithmic,booktabs,array,wrapfig,placeins,balance,enumerate}
\usepackage{graphicx}
\usepackage{textcomp}
\usepackage{hyperref}
\hypersetup{colorlinks,allcolors=black}
\usepackage{subfig}
\usepackage{caption}
\usepackage{adjustbox}
\usepackage{xcolor}
\usepackage{array}
\def\BibTeX{{\rm B\kern-.05em{\sc i\kern-.025em b}\kern-.08em
    T\kern-.1667em\lower.7ex\hbox{E}\kern-.125emX}}

    \makeatletter 
\newcommand{\linebreakand}{%
  \end{@IEEEauthorhalign}
  \hfill\mbox{}\par
  \mbox{}\hfill\begin{@IEEEauthorhalign}
}
\makeatother 
\begin{document}
\definecolor{myredcolor}{RGB}{254, 0, 0}
\definecolor{mygreencolor}{RGB}{0, 129, 0}
\definecolor{myyellowcolor}{RGB}{190, 191, 0}
\definecolor{mybluecolor}{RGB}{0, 1, 255}
\title{ Hybrid Quantum Neural Network based Indoor User Localization using Cloud Quantum Computing
\thanks{
$^{*}$ The first two authors contributed equally to the paper, and the names are arranged in alphabetical order.

This work was supported in part by the INSPIRE Faculty Fellowship from the Department of Science and Technology, Government of India (Reg. No.: IFA22-ENG 344) awarded to Neel Kanth Kundu and the New Faculty Seed
Grant (NFSG) from the Indian Institute of Technology Delhi.

This work has been accepted for presentation at the IEEE TENSYMP 2024 conference. Copyright may be transferred without notice, after which this version may no longer be accessible.

The simulation code for this paper is available at: \url{https://github.com/Sparshph1/Hybrid-Quantum-Neural-Network-based-User-Localization}
}}

\author{\IEEEauthorblockN{Sparsh Mittal$^{*}$}
\IEEEauthorblockA{\textit{Department of Physics} \\
\textit{Indian Institute of Technology Delhi}\\
New Delhi, India \\
ph1210215@physics.iitd.ac.in}
\and
\IEEEauthorblockN{Yash Chand$^{*}$}
\IEEEauthorblockA{\textit{Department of Physics} \\
\textit{   Indian Institute of Technology Delhi}\\
New Delhi, India  \\
ph1210829@physics.iitd.ac.in}
\linebreakand 
\IEEEauthorblockN{Neel Kanth Kundu}
\IEEEauthorblockA{\textit{Centre for Applied Research in Electronics (CARE)} \\ \textit{Bharti School of Telecommunication Technology and Management } \\
\textit{Indian Institute of Technology Delhi}\\
New Delhi, India  \\
\textit{Department of Electrical and Electronic Engineering} \\
\textit{University of Melbourne}  \\
Melbourne, VIC, Australia  \\
neelkanth@iitd.ac.in}
}

 \IEEEoverridecommandlockouts


\maketitle
\pagestyle{plain}
\begin{abstract}
This paper proposes a hybrid quantum neural network (HQNN) for indoor user localization using received signal strength indicator (RSSI) values. We use publicly available RSSI datasets for indoor localization using WiFi, Bluetooth, and Zigbee to test the performance of the proposed HQNN. We also compare the performance of the HQNN with the recently proposed quantum fingerprinting-based user localization method. Our results show that the proposed HQNN performs better than the quantum fingerprinting algorithm since the HQNN has trainable parameters in the quantum circuits, whereas the quantum fingerprinting algorithm uses a fixed quantum circuit to calculate the similarity between the test data point and the fingerprint dataset. Unlike prior works, we also test the performance of the HQNN and quantum fingerprint algorithm on a real IBM quantum computer using cloud quantum computing services. Therefore, this paper examines the performance of the HQNN on noisy intermediate scale (NISQ) quantum devices using real-world RSSI localization datasets. The novelty of our approach lies in the use of simple feature maps and ansatz with fewer neurons, alongside testing on actual quantum hardware using real-world data, demonstrating practical applicability in real-world scenarios.

\end{abstract}
\begin{IEEEkeywords}
quantum neural network, indoor localization, RSSI, quantum computing, IoT
\end{IEEEkeywords}

\section{Introduction}
\label{sec:intro}

Indoor localization is the process of determining or predicting the location of an object in an indoor environment. Traditional global navigation satellite systems (GNSS), such as GPS, face significant challenges in indoor settings due to signal attenuation and multipath effects caused by walls, ceilings, and other obstacles \cite{pahlavan2015localization}. To overcome these limitations, wireless localization techniques using WiFi, Bluetooth, and other wireless technologies have received widespread attention in recent years\cite{singh2021machine}\cite{gu2015energy}. The main attraction of these wireless techniques lies in their ability to predict location with fairly good accuracy without requiring additional infrastructure, leveraging the existing wireless network infrastructure already deployed in many indoor environments.

Wireless-signal-based indoor localization can broadly be classified into two main categories: geometric methods and fingerprinting methods. Geometric methods include triangulation, multilateration, and trilateration. These methods rely on measuring various parameters such as Time of Arrival (ToA), Time of Flight (ToF), and Angle of Arrival (AoA) to calculate the position of the target. While these methods are theoretically sound, they often suffer from practical limitations such as multipath distortion and non-line-of-sight (NLoS) issues, which degrade their performance in indoor environments. Additionally, these methods require high communication overhead and precise synchronization between units, further increasing the system's overall cost and complexity\cite{singh2021machine}.

Fingerprinting methods, on the other hand, involve creating a database of signal characteristics (such as Received Signal Strength Indicator (RSSI) values) at known locations during a training phase. During the online localization phase, the current signal measurements are compared with the stored fingerprints to determine the location. Traditional machine learning methods like K-Nearest neighbors (KNN)\cite{sadowski2020memoryless} and deep neural networks (DNN)\cite{guler2023quantum}\cite{eberechukwu2023fingerprinting} have been widely studied for RSSI fingerprinting-based indoor localization. However, these methods often struggle with issues related to dataset size, parameter tuning, and high computational training time, which can limit their accuracy and practicality in real-world applications\cite{singh2021machine}.

There has been a growing interest in exploring quantum computing for indoor localization in recent years \cite{zook2023quantum,urgelles2022application}. Quantum computing offers the potential for significant speedups in processing and optimization tasks compared to classical computing \cite{duong2022QML,duong2022QC}. Quantum algorithms, such as quantum fingerprinting, have been proposed for indoor localization tasks \cite{shokry2023quantum,shokry2022quantum,shokry2022qloc,shokry2022device}. However, these algorithms often lack trainable parameters, making them less suitable for handling high-interference and sparse data environments typically encountered in indoor localization scenarios\cite{spachos2020rssi}. Moreover, the number of qubits required increases with the dataset size, posing scalability challenges\cite{zook2023quantum}.

To address these challenges, we propose an HQNN for user localization using RSSI values, which is a common metric available in most WiFi devices. RSSI measures the path loss of wireless signals relative to a given distance, and this measurement can be derived using the Log-normal Distance Path Loss model \cite{seidel1992914}. Our approach combines the strengths of classical machine learning with the potential advantages of quantum computing, aiming to improve localization accuracy and efficiency. The main contributions of this paper are as follows:

\begin{enumerate}
    \item We propose an HQNN for user localization using RSSI values, which leverages the benefits of both classical and quantum computing paradigms.
    \item We investigate the performance of the proposed HQNN using publicly available RSSI datasets for indoor localization using WiFi, Bluetooth, and Zigbee, demonstrating its effectiveness across different wireless technologies and scenarios.
    \item We test the performance of the HQNN and quantum fingerprint algorithm on a real IBM quantum computer using cloud quantum computing services, providing practical insights into the feasibility and advantages of noisy intermediate scale quantum (NISQ) computing for indoor localization.
\end{enumerate}


In most cases, we find that HQNN converges more quickly than Classical NN during the training phase. As the interference in the RSSI dataset increases, the test performance of HQNN approaches that of the classical NN and eventually outperforms it. Furthermore, our results reveal that the performance of HQNN is significantly better than that of the quantum fingerprinting algorithm, specifically for high-interference scenarios \cite{eberechukwu2023fingerprinting}. One encouraging feature of HQNN is that its test performance on real IBM quantum hardware is not significantly lower than that of the simulator, suggesting that the model is not very sensitive to quantum hardware noise. Since HQNN has trainable parameters and a constant number of qubits, it is far more practical than quantum fingerprinting\cite{eberechukwu2023fingerprinting} and is more appropriate for high-interference data.

\section{Hybrid Quantum Neural Network}
\label{sec:HQNN}

The proposed HQNN integrates both quantum and classical neural network components to leverage the strengths of quantum computing in feature representation and classical neural networks in parameter optimization. Quantum computing, with its foundation in the principles of superposition and entanglement, offers a novel approach to processing information in ways that classical computing cannot achieve. The quantum part of HQNN is particularly effective in mapping the data to non-linear higher-dimensional Hilbert spaces, which allows for the exploration of complex data structures and relationships that are challenging for classical neural networks to model. By representing data in these high-dimensional spaces, the quantum component can capture intricate patterns and dependencies, enhancing the network's ability to learn from complex and high-dimensional datasets.

Scalability is another important factor in any quantum algorithm which could have practical implications in real-world applications. HQNN is a highly scalable architecture since the required number of qubits is equal to the dimension of the data point, which in our case is 3, due to the usage of three transmitters for the measurement of RSSI. The number of qubits is independent of the sample size of the dataset. Therefore, with the improvements in NISQ hardware, training and testing of a large dataset may be possible using HQNN, unlike quantum fingerprinting, where qubit requirement scales with the sample size of the dataset \cite{zook2023quantum}.

Furthermore, quantum entanglement, a unique property of quantum systems, enables the representation of intricate correlations between data points, which can be beneficial for modeling complex interactions in the data. These capabilities are particularly advantageous for applications involving high-dimensional and non-linear data, such as indoor localization, where traditional methods often struggle to achieve high accuracy. The integration of quantum computing into neural networks allows HQNNs to process complex data structures more naturally and efficiently, potentially leading to significant improvements in performance.

\begin{figure*}[ht]
    \centering
\includegraphics[width=0.8\textwidth]{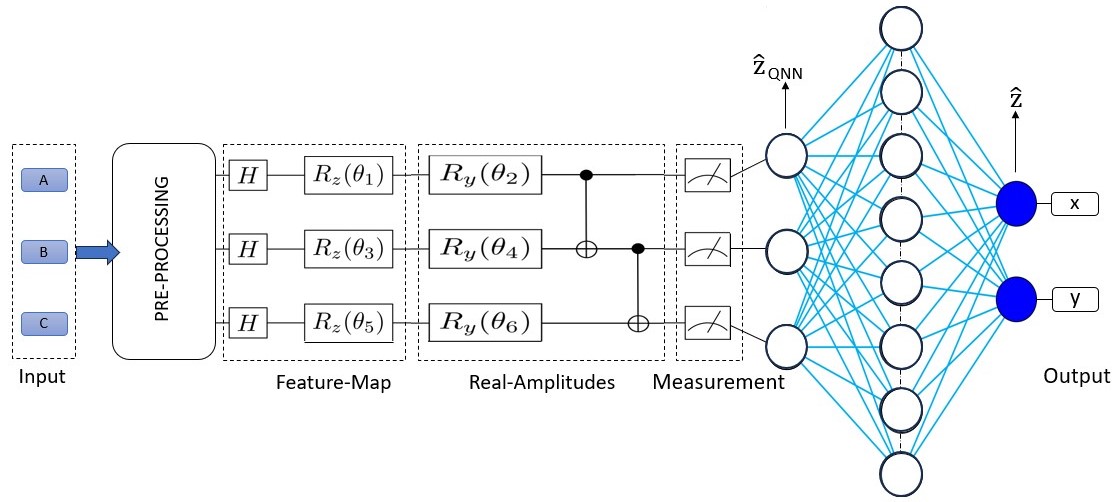}
    \caption{Schematic of the Hybrid Quantum-Classical Neural Network (HQNN)}
    \label{fig1}
\end{figure*}

On the other hand, the classical neural network component excels in parameter optimization through well-established training algorithms such as backpropagation and gradient descent. This component fine-tunes the parameters of the network to minimize prediction errors, leveraging the robust feature representations generated by the quantum part. The hybrid approach of HQNNs aims to combine the best of both worlds: the powerful feature extraction capabilities of quantum computing and the efficient parameter optimization techniques of classical neural networks. This synergy enables HQNNs to address challenging problems more effectively than purely classical or purely quantum approaches, paving the way for advancements in fields that require sophisticated data analysis and prediction capabilities. A schematic of the proposed HQNN is shown in Fig.\ \ref{fig1}. The different components of HQNN in Fig. \ref{fig1} are explained in the sub-sections below.

\subsection{Dataset Preparation}
We utilize publicly available RSSI datasets for indoor localization using WiFi, Bluetooth, and Zigbee \cite{spachos2020rssi}. RSSI measures the power present in a received radio signal, typically used to estimate the distance between the signal source and the receiver. In our study, the datasets comprise RSSI values collected from multiple access points or beacons strategically placed within an indoor environment. These RSSI values, which vary based on the proximity to the access points and the physical obstructions within the space, serve as the input features for our HQNN model.

The datasets include comprehensive RSSI measurements from three different sources: WiFi, Bluetooth, and Zigbee. Each dataset contains records where RSSI values are captured from multiple transmitters, providing a rich set of features that reflect the signal strength variations in different parts of the indoor environment. These RSSI values are accompanied by corresponding $(x, y)$ coordinates, which represent the actual physical location within the indoor space. These coordinates serve as the target outputs for the localization task.

For the purpose of training and evaluating our model, the datasets are divided into training and testing sets. The training set is used to train the neural networks, enabling them to learn the mapping from RSSI values to spatial coordinates. The testing set is reserved for evaluating the performance of the trained model, ensuring that it can generalize well to new, unseen data.

To facilitate processing by the neural networks, the RSSI values and corresponding coordinates are converted into appropriate tensor formats. Tensors, which are multi-dimensional arrays, are the standard data structures used in machine learning frameworks for efficient computation and manipulation of data. 

\subsection{Quantum Neural Network}
The quantum neural network (QNN) component of the HQNN is meticulously designed using the advanced quantum circuits provided by IBM Qiskit \cite{qiskit2024} to leverage the computational power of quantum mechanics. Central to this design is the use of a feature map, specifically the \texttt{ZZFeatureMap}, which plays a critical role in mapping the classical input data into a higher-dimensional quantum feature space using angle encoding. This transformation is essential for capturing complex data structures and relationships that are challenging to model with classical techniques alone. To parameterize the quantum circuit, an ansatz based on the \texttt{RealAmplitudes} circuit is employed. This ansatz introduces parameters that are optimized during the training phase, enabling the quantum circuit to adapt and learn from the input data effectively. It also contains linear entangling gates used for entangling the data from three qubits (see Fig.\ \ref{fig1}), which gives the advantage of quantum parallelism, a unique feature of quantum computing not available in the classical computing framework.

Given the substantial time requirements associated with the training of quantum circuits, especially when multiple layers are involved, the design strategically incorporates only a single layer of the quantum neural network. This decision balances the need for computational efficiency with the advantages offered by quantum feature mapping. The output of the quantum circuit is computed using the \texttt{EstimatorQNN} from Qiskit Machine Learning, a specialized tool designed to handle the intricacies of quantum neural network computations. These outputs are then integrated as part of the input to the overall hybrid model, allowing the combined system to benefit from both quantum and classical computational strengths. This integration ensures that the HQNN can process and learn from data in a highly efficient and sophisticated manner, paving the way for enhanced performance in complex applications.

\subsection{Classical Layers}
The classical layer (CL) component of HQNN consists of two fully connected layers, as shown in Fig.\ \ref{fig1}. The first layer transforms the input features into a higher-dimensional space using ReLU activation functions, while the second layer maps these transformed features to the output space using linear activation. This network is responsible for learning and optimizing the relationships between the input RSSI values and the target coordinates. Our experiments reveal that 32 neurons in the first layer are optimum.

\subsection{Integration and Training of HQNN}
The HQNN architecture is designed to merge the computational strengths of both quantum and classical layers. As shown in Fig. \ref{fig1}, this model takes 3D input data, which undergoes initial processing through a QNN. The QNN transforms the input data into an intermediate representation, denoted as $ \hat{\mathbf{z}}_{\rm QNN} $ (see Fig. \ref{fig1}). This intermediate output, \( \hat{\mathbf{z}}_{\rm QNN} \), is subsequently processed by the classical layers of the HQNN. The classical part of HQNN consists of a hidden layer with 32 neurons, which further processes the QNN output to produce the final prediction, \( \hat{\mathbf{z}} \), a 2D vector representing the predicted coordinates as shown in Fig. \ref{fig1}.

The training process of this combined model leverages gradient-based optimization techniques. Specifically, the loss function used to guide the training is the mean squared error (MSE) defined in  eq. (\ref{eq2}). The MSE is computed between the predicted coordinates, \( \hat{\mathbf{z}} \), and the true coordinates, \( \mathbf{z} \). Here, \( \mathbf{z} \) represents the actual 2D vector of final coordinates. The model is trained iteratively over multiple epochs to minimize the MSE, ensuring the predictions become progressively accurate.

During the training phase, it is critical to use a simulator due to the impracticality of running extensive training epochs directly on quantum hardware. Therefore, the IBM cloud simulator is employed to train the HQNN. The simulator provides a feasible environment for conducting numerous epochs of training. Once the training is complete, the model's performance is evaluated on a test dataset using IBM Quantum hardware. This step is essential to verify the accuracy and generalization capability of the HQNN in a real NISQ quantum computing environment.

The workflow of HQNN for indoor user localization can be described with the following steps:
\begin{enumerate}
\item  \textbf{Quantum Neural Network Processing}:
   The 3D input data \( \mathbf{x} \) is processed by the QNN to produce an intermediate output \( \hat{\mathbf{z}}_{\rm QNN} \).

\item  \textbf{Classical Layer Processing}:
   The intermediate output \( \hat{\mathbf{z}}_{\rm QNN} \) is fed into the classical layer with 32 neurons in the hidden layer. The hidden layer processes this to produce the final 2D output \( \hat{\mathbf{z}} \).
   

\item  \textbf{Combination and Prediction}:
   The final output of the HQNN is given by the output of the classical layers after processing the QNN output as
  \begin{equation}
      \hat{\mathbf{z}} = \text{CL}(\hat{\mathbf{z}}_{\rm QNN})
      \label{eq1}
  \end{equation}
  where $\text{CL} (\cdot) $ denotes the transformation produced by the classical layers of the HQNN.

\item  \textbf{Loss Function}:
   The loss function used for training the HQNN is the MSE between the predicted coordinates \( \hat{\mathbf{z}} \) and the true coordinates \( \mathbf{z} \):
   \begin{equation}
   L(\hat{\mathbf{z}}, \mathbf{z}) = {\frac{1}{n} \sum_{i=1}^{n} \left[ (\hat{x}_i - x_i)^2+ (\hat{y}_i - y_i)^2  \right]}
    \label{eq2}
  \end{equation}
   where \( n \) is the number of training data points.

\item  \textbf{Training and Optimization}:
During training, the parameters of the HQNN, including those of the QNN and the weights of the classical layers, are updated using a gradient-based optimization algorithm. The gradients of the loss function with respect to the model parameters are computed, and the parameters are updated iteratively to minimize the loss. This process is repeated over multiple epochs.

\begin{enumerate}[(a)]
    \item \textbf{Classical Gradient-Based Optimization (Adam)}:
For classical gradient-based optimization, such as the Adam optimizer, the update rule for the parameters $\theta$ of the model using the gradient $\nabla_\theta \mathcal{L}$ of the loss function $\mathcal{L}$ with respect to $\theta$ can be represented as:
\begin{equation}
\theta_{t+1} = \theta_{t} - \eta \cdot \text{AdamUpdate}(\nabla_\theta \mathcal{L})
\end{equation}
where $\eta$ is the learning rate, $t$ is the iteration number, and AdamUpdate represents the Adam optimizer update step \cite{kingma2014adam}. The Adam optimizer combines the advantages of two other popular optimization techniques: AdaGrad and RMSProp. It adapts the learning rate for each parameter by computing individual adaptive learning rates.

The Adam update rule can be detailed as follows:
\begin{align}
    m_t &= \beta_1 m_{t-1} + (1 - \beta_1) \nabla_\theta \mathcal{L}_t \\
    v_t &= \beta_2 v_{t-1} + (1 - \beta_2) (\nabla_\theta \mathcal{L}_t)^2 \\
    \hat{m}_t &= \frac{m_t}{1 - \beta_1^t} \\
    \hat{v}_t &= \frac{v_t}{1 - \beta_2^t} \\
    \theta_{t+1} &= \theta_t - \eta \frac{\hat{m}_t}{\sqrt{\hat{v}_t} + \epsilon}
\end{align}
where
 $m_t$ and $v_t$ are the first and second-moment estimates, respectively, and $\beta_1$ and $\beta_2$ are the exponential decay rates for the moment estimates, typically set to 0.9 and 0.999, respectively. Further,
 $\hat{m}_t$ and $\hat{v}_t$ are the bias-corrected moment estimates and
 $\epsilon$ is a small constant (e.g., $10^{-8}$) to prevent division by zero.
For more details on the Adam optimization algorithm, see \cite{kingma2014adam}.

    \item \textbf{Quantum Gradient Based Optimization (Parameter Shift Rule)}:
    For quantum gradient-based optimization using the parameter-shift rule, the update rule for the parameters $\Phi$ of the quantum model is given by
    \begin{equation}
    \Phi_{t+1} = \Phi_{t} - \eta \cdot \Delta \Phi
    \end{equation}
    where $\eta$ is the learning rate, $t$ is the iteration number, and $\Delta \Phi$ is the parameter shift calculated as:
    \begin{equation}
    \Delta \Phi = \frac{1}{2} \left( \mathcal{L}(\Phi + \frac{\pi}{2}) - \mathcal{L}(\Phi - \frac{\pi}{2}) \right)
    \end{equation}
    This approach uses the parameter-shift rule to estimate the gradient of the loss function $\mathcal{L}$ with respect to $\Phi$.
\end{enumerate}
\item \textbf{Performance Evaluation}:
   The performance of the HQNN is evaluated using the RMSE of the predictions on the test dataset run on an IBM quantum simulator as well as on real IBM Quantum hardware (ibm\_kyiv and ibm\_nazca).
   \begin{equation}
   \text{RMSE} = \sqrt{{\frac{1}{n_{\rm test}} \sum_{i=1}^{n_{\rm test}} \left[ (\hat{x}_i - x_i)^2+ (\hat{y}_i - y_i)^2  \right]}}
   \label{eqn11}
   \end{equation}
   where $n_{\rm test}$ is the number of testing data points.
\end{enumerate}


In summary, the integration and training of the HQNN involve processing 3D input data through the QNN, feeding the QNN output to the classical layers with $32$ neurons, and iteratively optimizing the combined model using gradient-based methods to minimize the MSE between predicted and true coordinates as defined in (\ref{eq2}).


\subsection{Implementation on IBM Quantum Hardware}
The performance of the HQNN is evaluated on real IBM quantum hardware using cloud quantum computing services. Previous works on quantum computing-based localization have used only the simulator\cite{eberechukwu2023fingerprinting} to test the performance of the model, thereby not incorporating the effects of quantum hardware noise. The quantum circuits are transpiled and optimized for the specific quantum backend (IBM\_kyiv and IBM\_nazca) to ensure efficient execution. The results obtained from the quantum hardware are compared with those from classical simulations to assess the impact of quantum noise and hardware-specific characteristics on the model's performance.


\section{Indoor Localization Datasets}
\label{sec:datasets}

The indoor localization dataset used in this study is real-world data that encompasses a wide range of indoor environments \cite{spachos2020rssi}. The localization data was collected in three different indoor scenarios, which are described in the subsections below. These datasets provide valuable insights into the performance of the proposed HQNN across different scenarios. All three scenarios contain data corresponding to Zigbee (IEEE $802.15.4$), Bluetooth Low Energy (BLE), and WiFi (IEEE $802.11$n $2.4$ GHz band) wireless technologies. 

\subsection{Small Meeting Rooms with Low Interference (Sc-1)}
The dataset is for a small meeting room of dimension $6  \times 5.5$ m with low interference, providing an ideal environment for evaluating localization algorithms under minimal disturbance. This dataset offers detailed measurements of RSSI values in a controlled setting, enabling precise analysis of localization accuracy. The training dataset is composed of 49 points and tested on 10 data points.

\subsection{Small Meeting Rooms with High Interference (Sc-2)}
This simulates environments with high interference deliberately introduced, mimicking real-world scenarios where electronic devices and other wireless equipment may cause signal degradation in a room of dimension 5.8 x 5.3 m. This dataset allows for the assessment of localization algorithms' robustness in challenging conditions. The training dataset is composed of 16 points and tested on 6 data points.

\subsection{Large Computer Labs with Average Interference (Sc-3)}
Additionally, a dataset representing a large computer lab environment of dimension 10.8 x 7.2 m is included, offering a more complex setting with multiple obstructions and dynamic interference sources like moving individuals and active wireless devices. This dataset presents a realistic indoor environment with varying levels of interference and obstruction, facilitating a comprehensive evaluation of the proposed HQNN. The training dataset is composed of 40 points and tested on 16 data points.

\section{Numerical Results}
\label{sec:results}
The model was trained using multiple learning rates, different optimizers, and varying numbers of neurons in the hidden classical layer to determine the optimal configuration. Through extensive experimentation, it was found that the Adam optimizer consistently produced the best results compared to other optimization techniques. Specifically, the hidden classical layer with $32$ neurons and a learning rate of $0.001$ emerged as the ideal setup. This configuration was carefully tuned for both training and testing data, ensuring that the model avoided both over-fitting and under-fitting. MSE is used as the loss function for training, and RMSE is used to report accuracy on test data for dimensional consistency of distance (in m). 

\subsection{Experimental Setup and Parameters}

\subsubsection*{Optimizers}
Various optimizers, including NAdam, SGD, Adagrad, and Adam, were tested. Adam stood out due to its adaptive learning rate capabilities and efficient handling of sparse gradients, leading to faster convergence and better overall performance.

\subsubsection*{Neurons in Hidden Layer}
A hidden layer with $32$ neurons was identified as optimal. This balance of complexity allowed the model to learn effectively without becoming too computationally expensive or prone to overfitting.

\subsubsection*{Learning Rate}
A learning rate of $0.0001$ was found to be ideal. This low learning rate ensured stable convergence and fine-tuning of the model’s weights.

\subsection{Comparison with Classical Neural Networks (NN), K-Nearest Neighbors (KNN) and Quantum Fingerprinting }
\begin{figure*}[h]
    \centering
    \begin{minipage}[b]{0.85\linewidth}
    
    \subfloat[SC1 Zigbee]{
        \includegraphics[width=0.3\textwidth]{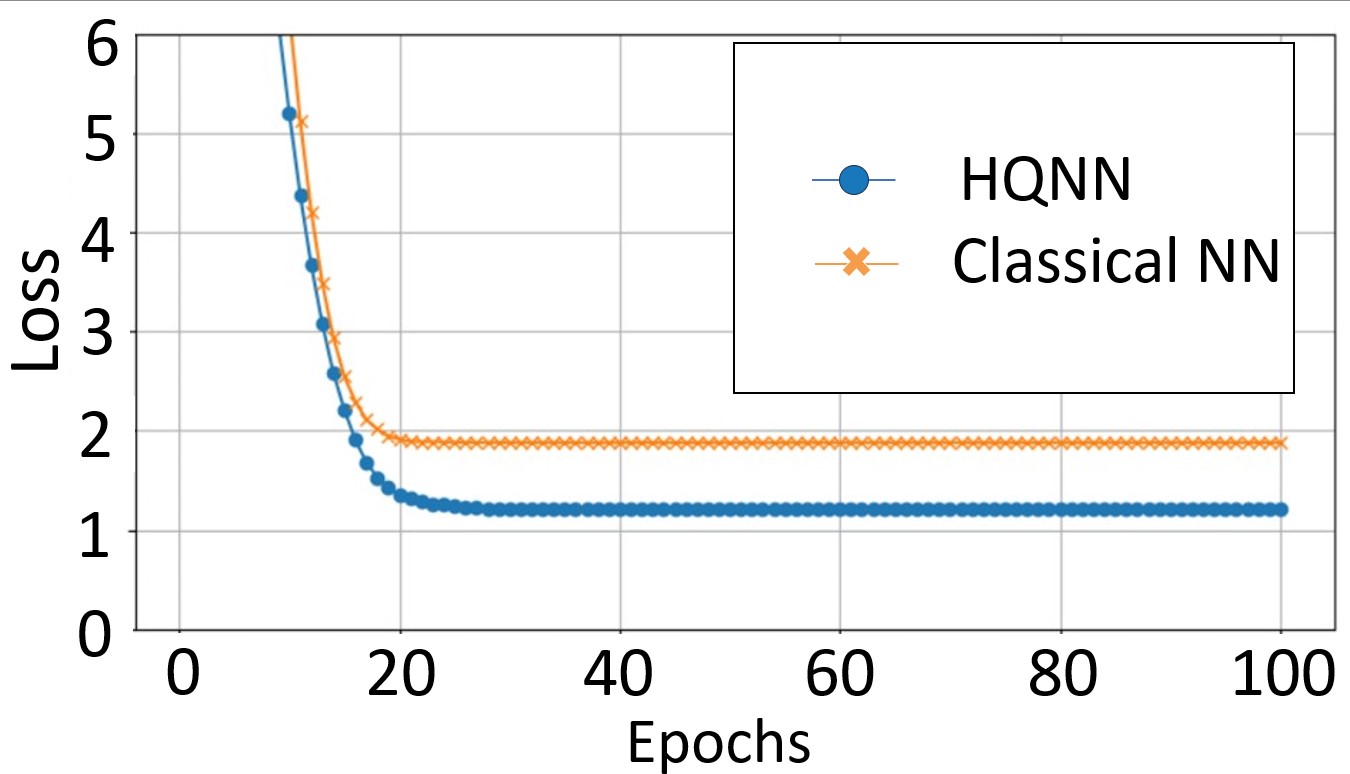}
        \label{fig:SC_1_Zigbee}
    }
    \hfill
    \subfloat[SC1 Bluetooth]{
        \includegraphics[width=0.3\textwidth]{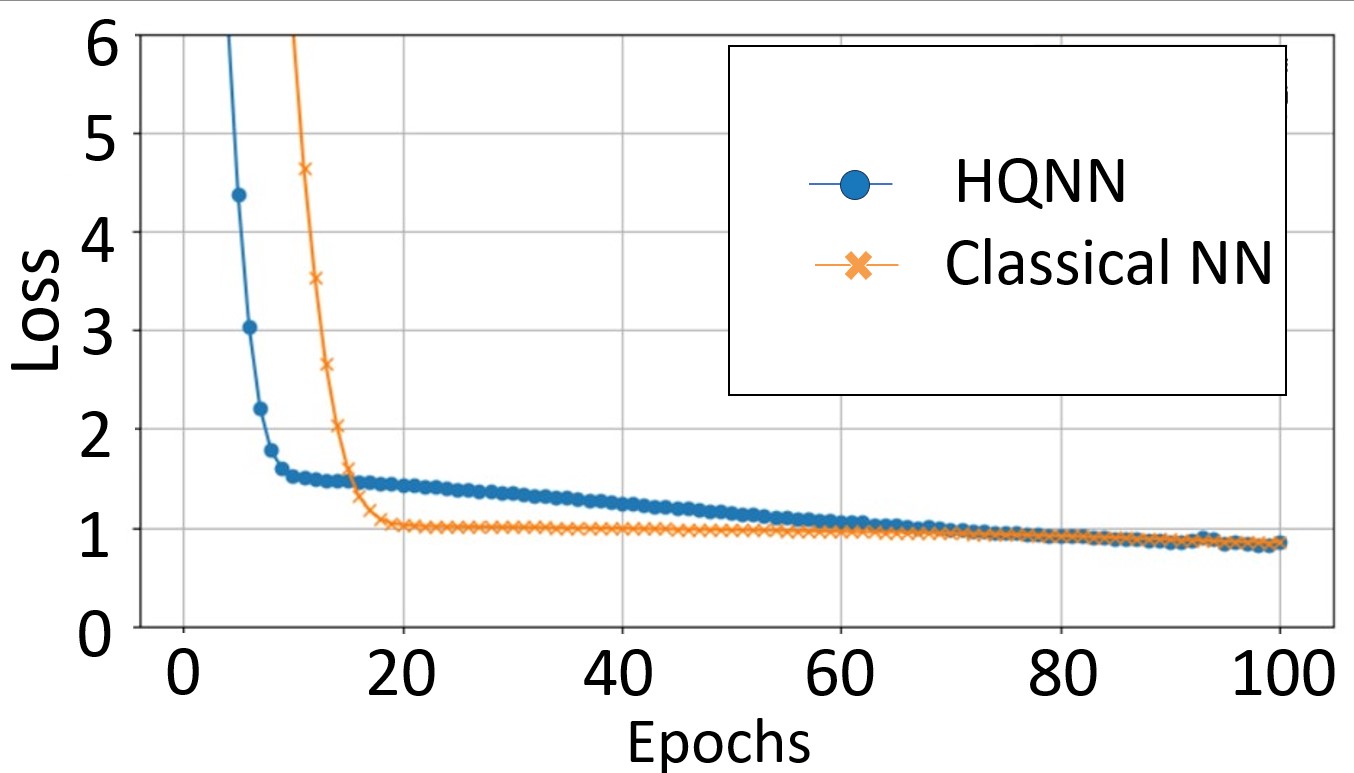}
        \label{fig:SC_1_Bluetooth}
    }
    \hfill
    \subfloat[SC1 WiFi]{
        \includegraphics[width=0.3\textwidth]{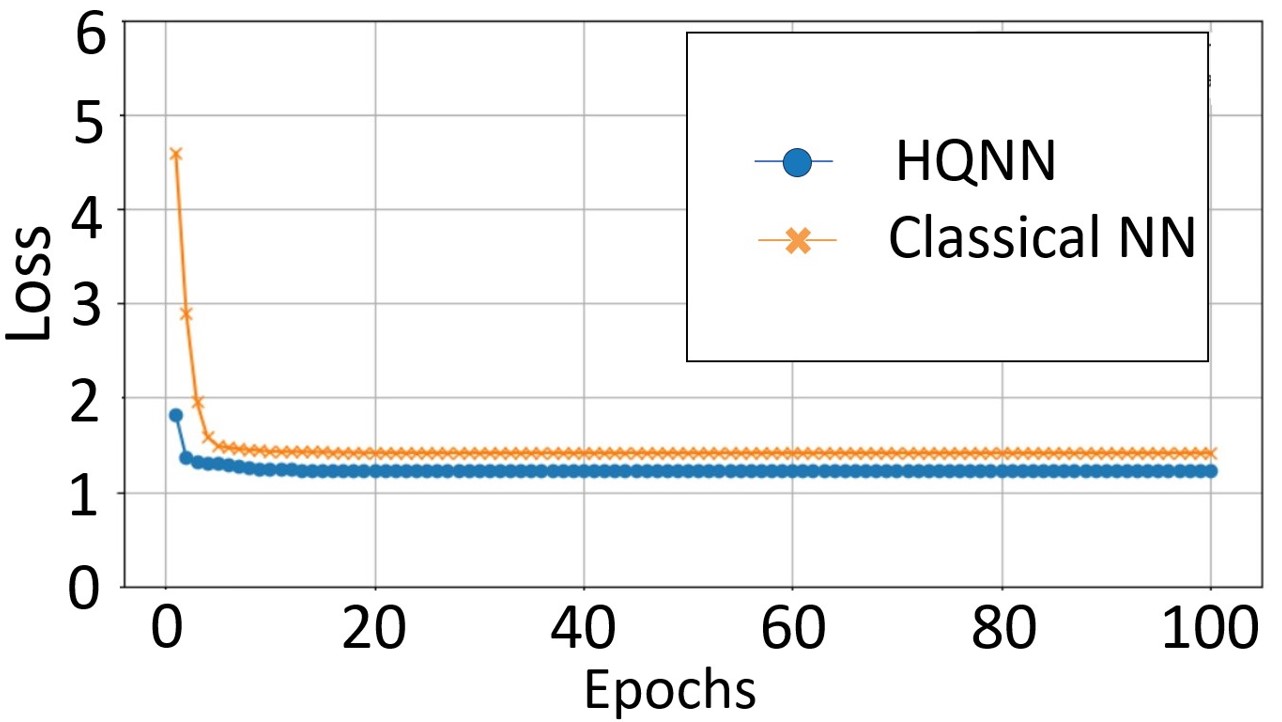}
        \label{fig:SC_1_WiFi}
    }
    
    \vspace{0.2cm}
    
    \subfloat[SC2 Zigbee]{
        \includegraphics[width=0.3\textwidth]{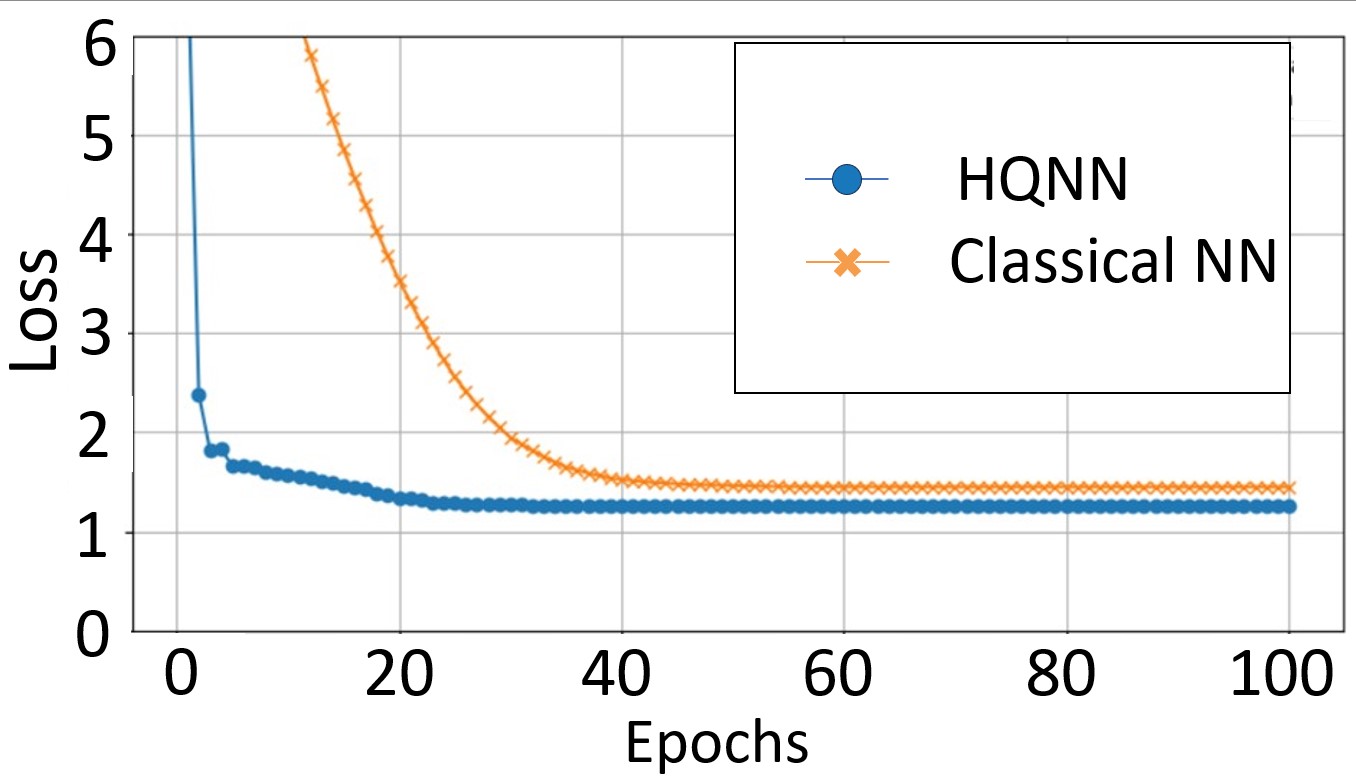}
        \label{fig:SC_2_Zigbee}
    }
    \hfill
    \subfloat[SC2 Bluetooth]{
        \includegraphics[width=0.3\textwidth]{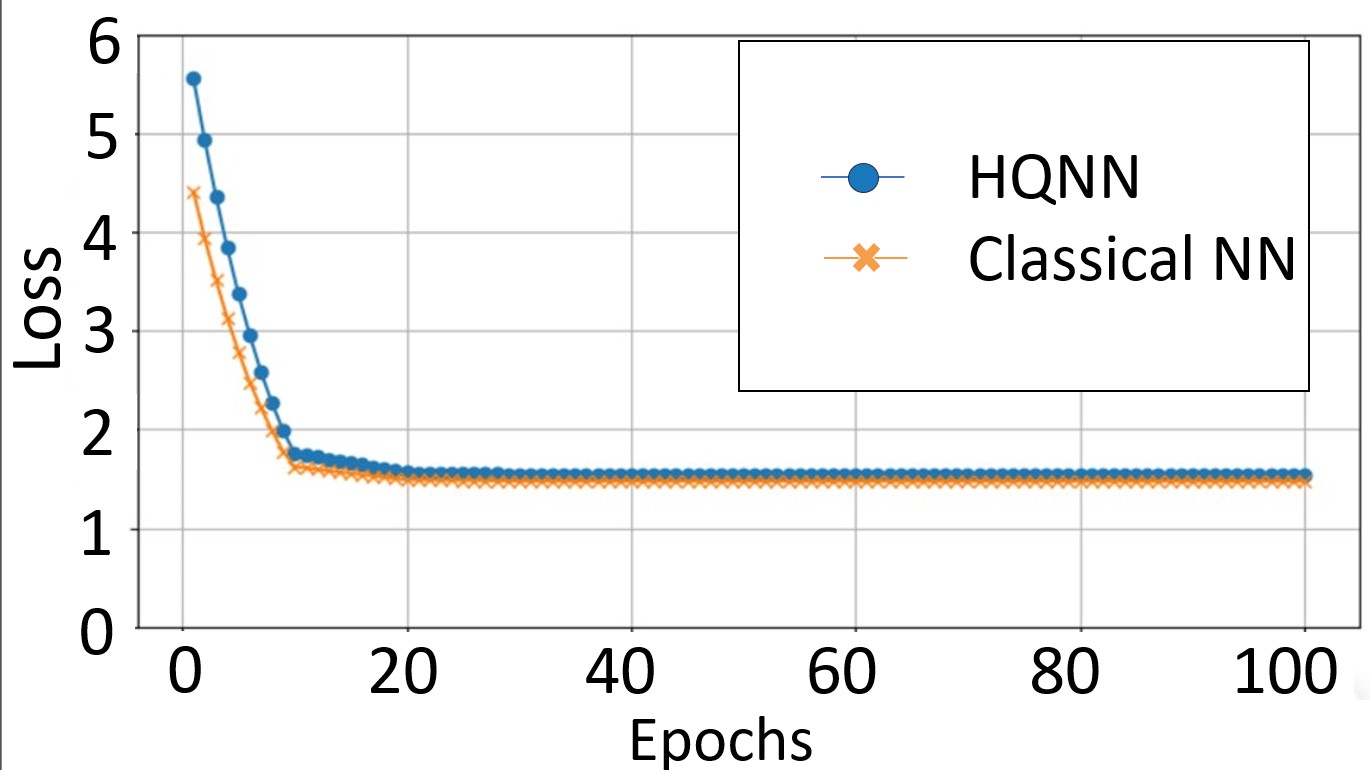}
        \label{fig:SC_2_Bluetooth}
    }
    \hfill
    \subfloat[SC2 WiFi]{
        \includegraphics[width=0.3\textwidth]{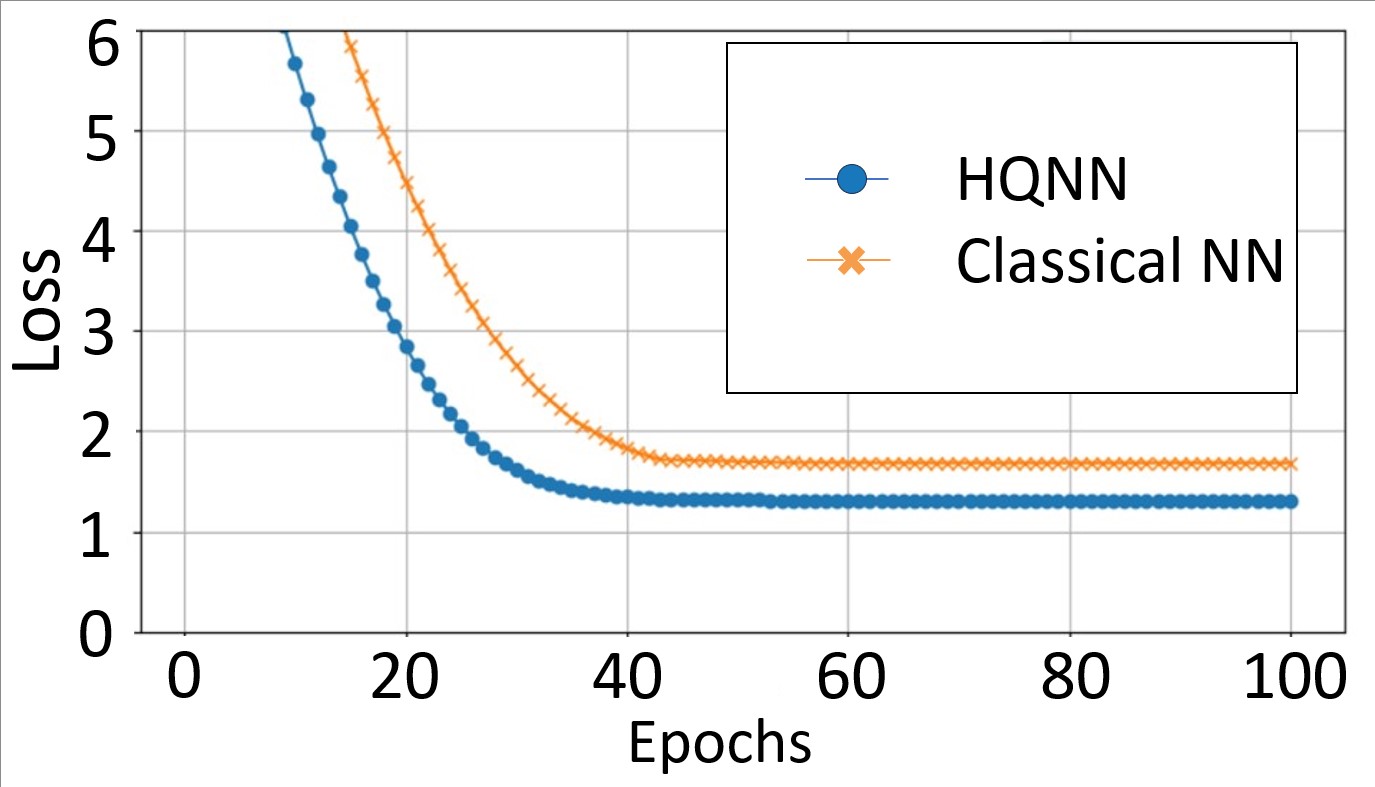}
        \label{fig:SC_2_WiFi}
    }
    
    \vspace{0.2cm}
    
    \subfloat[SC3 Zigbee]{
        \includegraphics[width=0.3\textwidth]{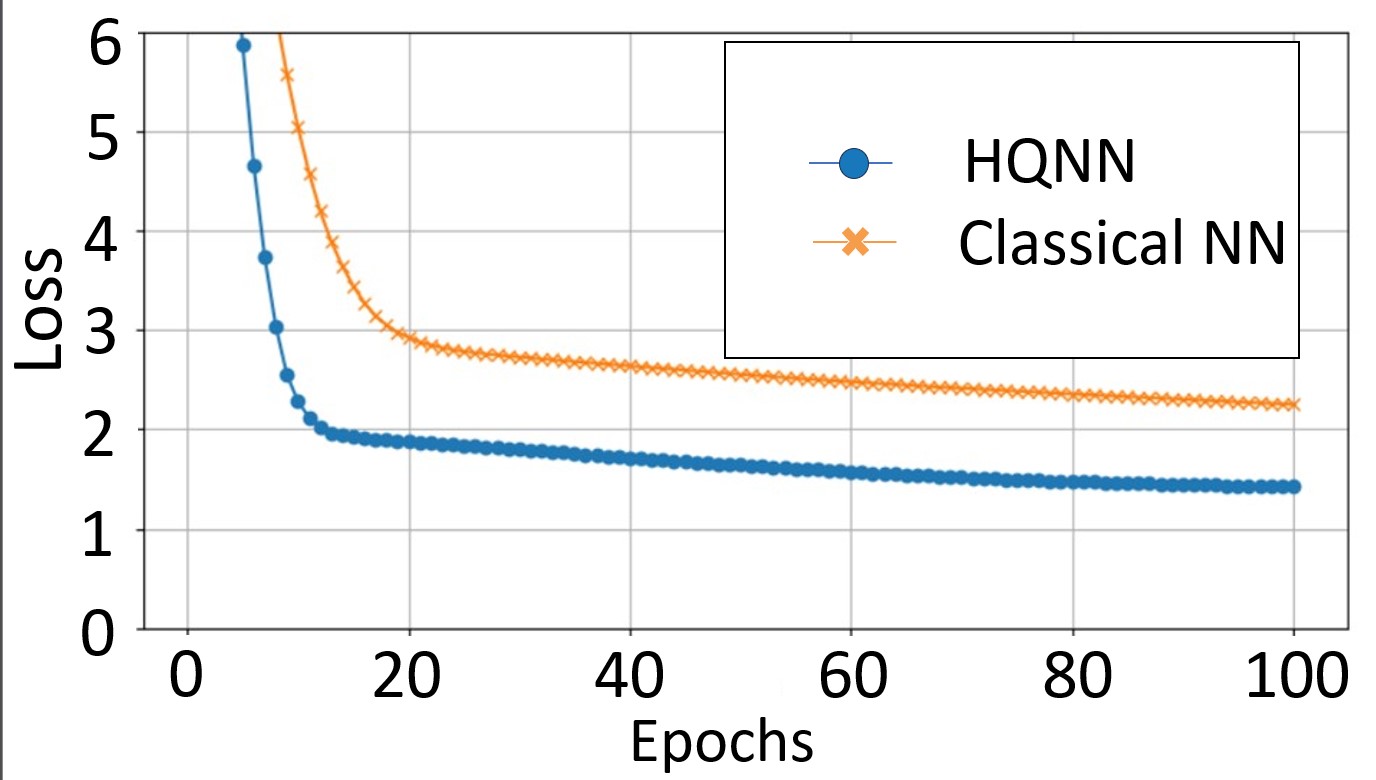}
        \label{fig:SC_3_Zigbee}
    }
    \hfill
    \subfloat[SC3 Bluetooth]{
        \includegraphics[width=0.3\textwidth]{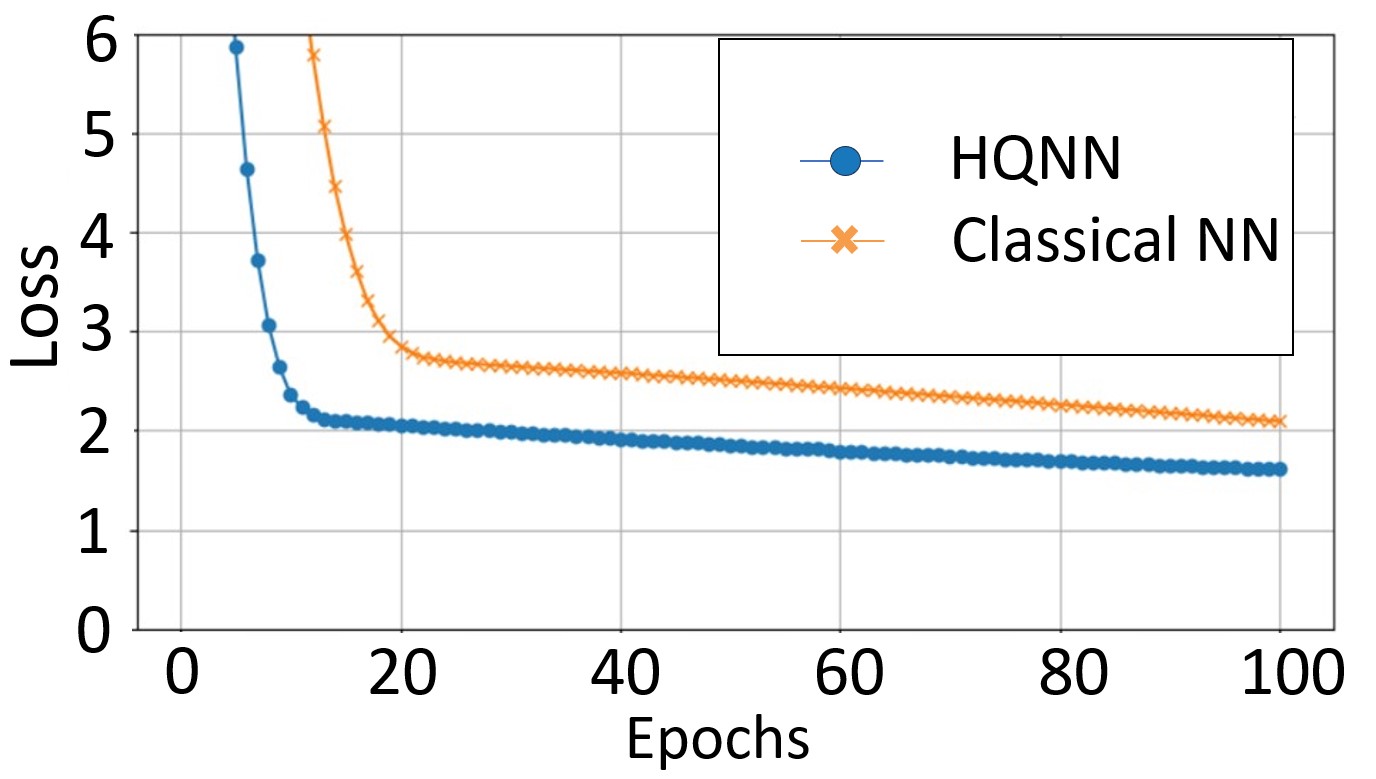}
        \label{fig:SC_3_Bluetooth}
    }
    \hfill
    \subfloat[SC3 WiFi]{
        \includegraphics[width=0.3\textwidth]{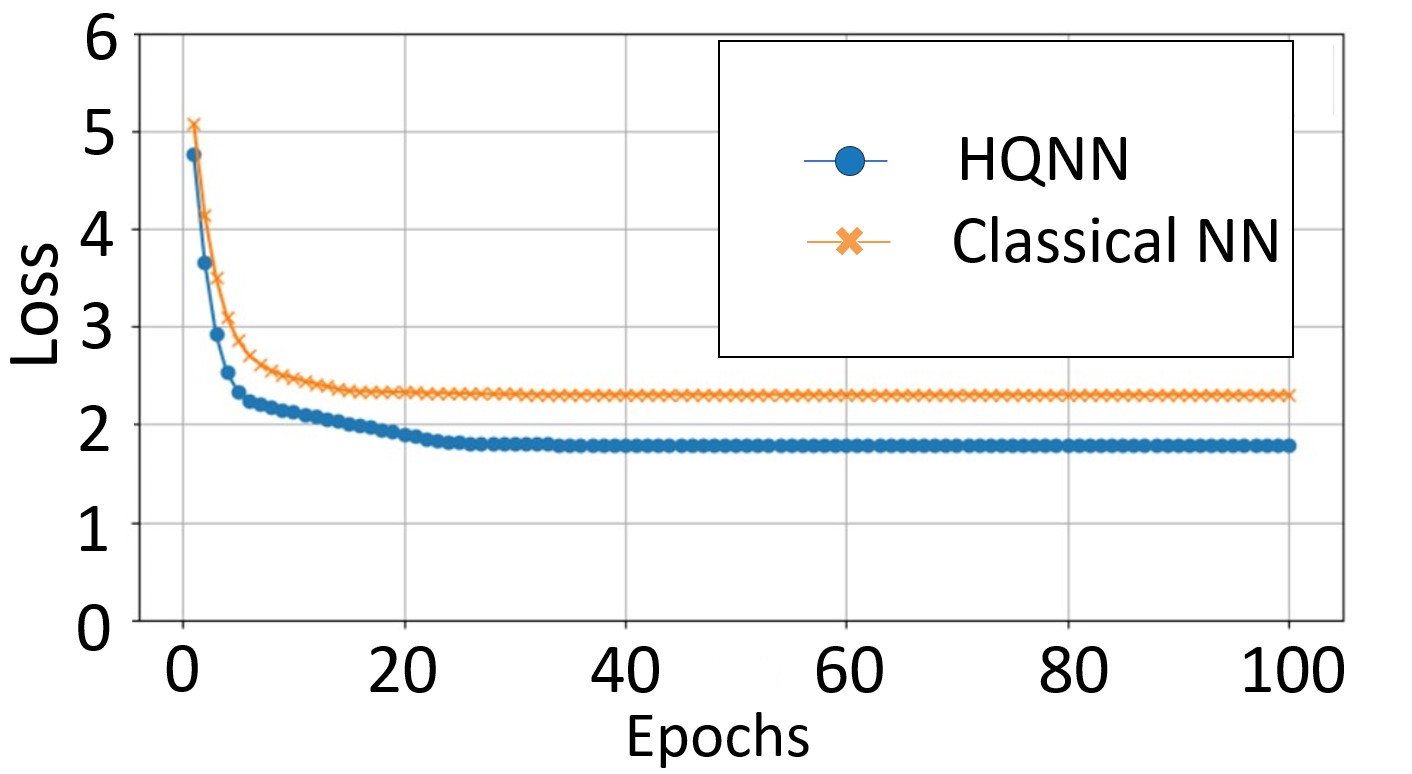}
        \label{fig:SC_3_WiFi}
    }

    \caption{Training Loss vs Epochs for HQNN and classical NN for different scenarios and devices}
    \label{fig:graphs}
    \end{minipage}
\end{figure*}

\begin{table*}[htp]
\centering
\caption{Comparison of RMSE for Various Scenarios and Algorithms}
\begin{tabular}{|l|c|c|c|c|c|}
\hline
\textbf{Device} & \textbf{NN} & \textbf{KNN } & \textbf{Quantum Fingerprint} & \textbf{HQNN (Hardware)} & \textbf{HQNN (Simulator)} \\
\hline
\textbf{Bluetooth Sc-1} & 1.277 & 1.220 & 1.300 & 1.273 & 1.216 \\
\textbf{Wifi Sc-1} & 1.290 & 1.296 & 1.526 & 1.367 & 1.289\\
\textbf{Zigbee Sc-1} & 1.368 & 1.342 & 1.471 & 1.196 & 1.108\\
\textbf{Bluetooth Sc-2} & 1.175 & 1.273 & 2.171 & 1.353 & 1.279\\
\textbf{Wifi Sc-2} & 1.583 & 1.272 & 1.755 & 1.339 & 1.231\\
\textbf{Zigbee Sc-2} & 0.819 & 1.261 & 2.105 & 0.971 & 0.918\\
\textbf{Bluetooth Sc-3} & 1.384 & 1.432 & 2.396 & 1.337 & 1.201\\
\textbf{Wifi Sc-3} & 0.999 & 1.201 & 2.532 & 1.224 & 1.138\\
\textbf{Zigbee Sc-3} & 1.306 & 1.379 & 2.616 & 1.471 & 1.298\\
\hline
\end{tabular}
\label{tab:results}
\end{table*}

\begin{figure*}[htp]
    \centering
    \begin{minipage}[b]{0.95\linewidth}
    
    \subfloat[Scenario 1]{
        \includegraphics[width=0.3\textwidth]{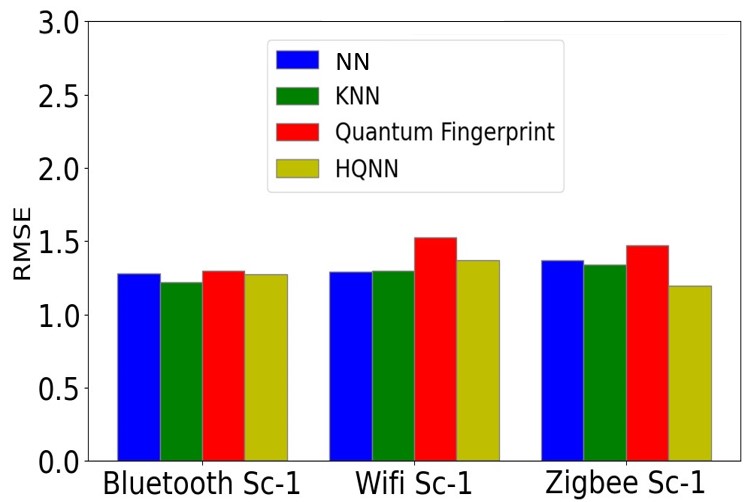}
        \label{Scenario 1}
    }
    \hfill
    \subfloat[Scenario 2]{
        \includegraphics[width=0.3\textwidth]{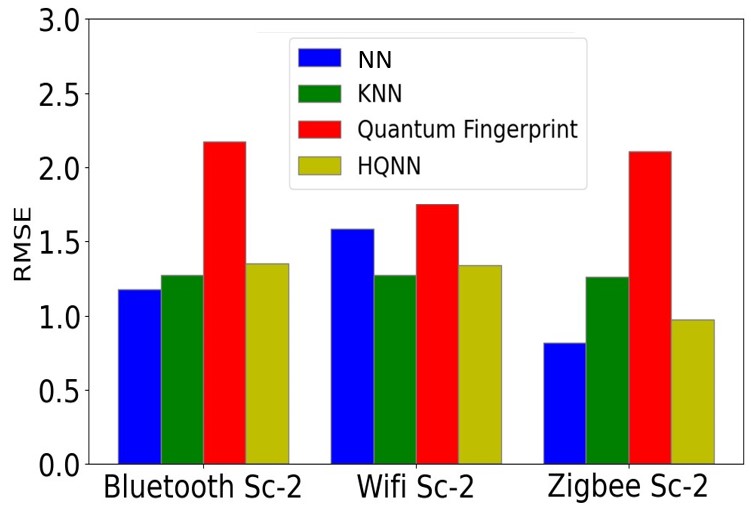}
        \label{Scenario 2}
    }
    \hfill
    \subfloat[Scenario 3]{
        \includegraphics[width=0.3\textwidth]{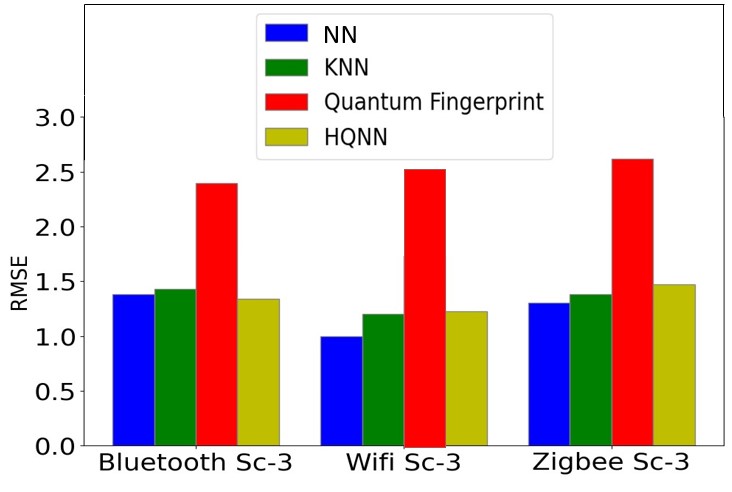}
        \label{Scenario 3}
    }
\end{minipage}
\caption{\centering RMSE comparison of test dataset on IBM quantum hardware for different scenarios:
{\textcolor{mybluecolor}{Classical NN}}, {\textcolor{mygreencolor}{KNN}}, {\textcolor{myredcolor}{Quantum Fingerprinting}}, and {\textcolor{myyellowcolor}{HQNN}}}
\label{fig:bar_graphs}
\end{figure*}
Classical NNs are powerful tools for a wide range of applications, but they have limitations when it comes to handling complex, interference-laden data. We use a classical NN with two hidden layers with $128$ and $64$ neurons, which are much more than $6$ parameters in the quantum hidden layer and $32$ neurons in the classical layer of HQNN as shown in Fig.\ \ref{fig1}. The classical NN model in our tests showed a notable performance drop in Scenario 3 while training. This can be attributed to its reliance on classical computation, which may not be as effective in capturing the subtleties and interactions within highly noisy data. In comparison, the HQNN leverages quantum principles to enhance data representation and processing. The quantum layer of the HQNN can process complex patterns that are challenging for classical models, leading to better performance in scenarios with significant interference. Comparable test performance can be attributed to noise-prone quantum hardware.

Fig.\ \ref{fig:graphs} shows the plots of the training loss versus epochs of the HQNN and Classical NN for all the scenarios. It can be observed that our proposed HQNN model achieves rapid convergence. Furthermore, it can be observed that HQNN achieves a lower value for loss function than the classical NN within a minimal epoch count for all the different cases.

Next, we compare the test performance of HQNN with K-Nearest Neighbors (KNN), which is a classical machine-learning model \cite{kramer2013k}, and the recently proposed quantum fingerprinting method \cite{zook2023quantum}. Comparison of the RMSE of different models, i.e., KNN, Quantum  Fingerprinting, classical NN, and the proposed HQNN, is shown in Table \ref{tab:results} and Fig. \ref{fig:bar_graphs}. The performance of HQNN is evaluated using the \textit{ibmq\_qasm\_simulator} and the real quantum device, \textit{ibm\_kyiv}. It can be observed that HQNN clearly excels in the quantum fingerprinting model and is comparable to other classical models. Scenario 3, characterized by heavy interference data, highlighted the advantages of the HQNN over classical NN. The classical NN struggled with the high level of complexity, failing to capture the intricate patterns within the data. In contrast, the HQNN demonstrated superior performance due to its quantum components' ability to process and represent complex data structures more effectively. For instance, in Scenario 3 with Zigbee, the HQNN (Simulator) achieved an RMSE of $1.298$ compared to $2.616$ with Quantum Fingerprinting and $1.306$ of classical NN. Similarly, for Bluetooth in Scenario 1, the HQNN (Simulator) showed the lowest RMSE of $1.216$ among all models. Across all scenarios and device types, the HQNN consistently performed better, particularly excelling in high-interference environments.

\subsection{Limitations of K-Nearest Neighbors \textnormal(KNN)}

The KNN algorithm is a simple yet effective non-parametric method used for classification and regression\cite{kramer2013k}. It excels in scenarios with clearly defined clusters and relatively low noise. However, KNN has its drawbacks:
\begin{enumerate}[(a)]
    
    \item \textbf{Scalability}: KNN suffers from scalability issues as the dataset grows, since it requires computing the distance to every point in the dataset for each prediction.
    \item \textbf{High-Dimensional Data}: In high-dimensional spaces, the concept of proximity becomes less meaningful, which can degrade the performance of KNN.
    \item \textbf{Interference Handling}: KNN does not perform well in the presence of high interference, as it relies on distance metrics that can be easily distorted by noise.
\end{enumerate}

In our experiments, KNN was less effective in Scenario 3 with heavy interference. The quantum component of the HQNN allowed it to outperform KNN by effectively managing the complexity and noise in the data, something that classical distance-based methods struggled with.

\subsection{Limitations of Quantum Fingerprinting}

Quantum fingerprinting is a method designed for efficient comparison of data strings \cite{shokry2023quantum}. It offers high efficiency with sub-quadratic time and space complexities, making it an excellent choice for tasks that involve low-disturbance data. However, it lacks training parameters, which means it does not adapt or learn from data in the way that supervised models do. While quantum fingerprinting excels in specific prediction tasks with low noise, it is not suitable for scenarios with average or high interference, as it cannot adjust to varying data patterns through training. In contrast, our proposed HQNN model can be trained to recognize and adapt to complex, noisy data, providing a significant advantage in such environments.




\section{Conclusions}
\label{sec:con}

In this paper, we have proposed a hybrid quantum neural network (HQNN) that combines the strengths of classical and quantum computing for indoor user localization. HQNN leverages quantum parallelism for faster, more intricate pattern analysis while maintaining robustness with classical neural networks. This approach overcomes challenges like vanishing gradients and overfitting in non-linear datasets. Experimental results show significant improvement in localization accuracy compared to existing algorithms.

While promising, HQNNs face certain challenges that must be overcome for widespread deployment of HQNN for user localization in future wireless networks. Training large datasets on current quantum hardware is time-consuming and challenging. Furthermore, effective quantum error correction is crucial to mitigate quantum noise and improve the performance of HQNN on real quantum hardware. Addressing these challenges is essential for the practical deployment of HQNNs. As we refine these models and overcome limitations, HQNNs have the potential to revolutionize indoor localization and become a cornerstone technology for intelligent IoT ecosystems.

\section*{Acknowledgment}
The authors acknowledge the use of advanced IBM~Quantum services provided by the IBM~Quantum Network Hub at the University of Melbourne.

\bibliographystyle{ieeetr}
\bibliography{Refs}

\begin{thebibliography}{10}

\bibitem{pahlavan2015localization}
K.~Pahlavan, P.~Krishnamurthy, and Y.~Geng, ``Localization challenges for the emergence of the smart world,'' {\em IEEE Access}, vol.~3, pp.~3058--3067, 2015.

\bibitem{singh2021machine}
N.~Singh, S.~Choe, and R.~Punmiya, ``Machine learning based indoor localization using {Wi-Fi RSSI} fingerprints: An overview,'' {\em IEEE Access}, vol.~9, pp.~127150--127174, 2021.

\bibitem{gu2015energy}
Y.~Gu and F.~Ren, ``Energy-efficient indoor localization of smart hand-held devices using bluetooth,'' {\em IEEE Access}, vol.~3, pp.~1450--1461, 2015.

\bibitem{sadowski2020memoryless}
S.~Sadowski, P.~Spachos, and K.~N. Plataniotis, ``Memoryless techniques and wireless technologies for indoor localization with the internet of things,'' {\em IEEE Internet of Things Journal}, vol.~7, no.~11, pp.~10996--11005, 2020.

\bibitem{guler2023quantum}
E.~Guler, M.~T. Kak{\i}z, T.~{\c{C}}avdar, F.~B. G{\"u}nay, and B.~{\c{S}}anal, ``A quantum machine learning approach for detecting user locations,'' {\em Journal of Millimeterwave Communication, Optimization and Modelling}, vol.~3, no.~1, pp.~1--8, 2023.

\bibitem{eberechukwu2023fingerprinting}
N.~P. Eberechukwu, M.~Jeong, H.~Park, S.~W. Choi, and S.~Kim, ``Fingerprinting-based indoor localization with hybrid quantum-deep neural network,'' {\em IEEE Access}, 2023.

\bibitem{zook2023quantum}
Y.~Zook, A.~Shokry, and M.~Youssef, ``A quantum fingerprinting algorithm for next generation cellular positioning,'' {\em IEEE Journal on Selected Areas in Communications}, 2023.

\bibitem{urgelles2022application}
H.~Urgelles, P.~Picazo-Mart{\'\i}nez, and J.~F. Monserrat, ``Application of quantum computing to accurate positioning in {6G} indoor scenarios,'' in {\em ICC 2022-IEEE International Conference on Communications}, pp.~643--647, IEEE, 2022.

\bibitem{duong2022QML}
T.~Q. Duong, J.~A. Ansere, B.~Narottama, V.~Sharma, O.~A. Dobre, and H.~Shin, ``Quantum-inspired machine learning for {6G}: Fundamentals, security, resource allocations, challenges, and future research directions,'' {\em IEEE Open Journal of Vehicular Technology}, vol.~3, pp.~375--387, 2022.

\bibitem{duong2022QC}
T.~Q. Duong, L.~D. Nguyen, B.~Narottama, J.~A. Ansere, D.~Van~Huynh, and H.~Shin, ``Quantum-inspired real-time optimization for 6g networks: Opportunities, challenges, and the road ahead,'' {\em IEEE Open Journal of the Communications Society}, vol.~3, pp.~1347--1359, 2022.

\bibitem{shokry2023quantum}
A.~Shokry and M.~Youssef, ``Quantum fingerprinting for heterogeneous devices localization,'' {\em Computer Communications}, vol.~204, pp.~43--51, 2023.

\bibitem{shokry2022quantum}
A.~Shokry and M.~Youssef, ``A quantum algorithm for {RF}-based fingerprinting localization systems,'' in {\em 2022 IEEE 47th Conference on Local Computer Networks (LCN)}, pp.~18--25, IEEE, 2022.

\bibitem{shokry2022qloc}
A.~Shokry and M.~Youssef, ``Qloc: A realistic quantum fingerprint-based algorithm for large scale localization,'' in {\em 2022 IEEE International Conference on Quantum Computing and Engineering (QCE)}, pp.~238--246, IEEE, 2022.

\bibitem{shokry2022device}
A.~Shokry and M.~Youssef, ``Device-independent quantum fingerprinting for large scale localization,'' in {\em 2022 20th Mediterranean Communication and Computer Networking Conference (MedComNet)}, pp.~208--215, IEEE, 2022.

\bibitem{spachos2020rssi}
P.~Spachos, ``{RSSI} dataset for indoor localization fingerprinting,'' {\em IEEE Dataport}, 2020.

\bibitem{seidel1992914}
S.~Y. Seidel and T.~S. Rappaport, ``914 {MHz} path loss prediction models for indoor wireless communications in multifloored buildings,'' {\em IEEE Transactions on Antennas and Propagation}, vol.~40, no.~2, pp.~207--217, 1992.

\bibitem{qiskit2024}
A.~Javadi-Abhari, M.~Treinish, K.~Krsulich, C.~J. Wood, J.~Lishman, J.~Gacon, S.~Martiel, P.~D. Nation, L.~S. Bishop, A.~W. Cross, B.~R. Johnson, and J.~M. Gambetta, ``Quantum computing with {Q}iskit,'' 2024.

\bibitem{kingma2014adam}
D.~P. Kingma and J.~Ba, ``Adam: A method for stochastic optimization,'' {\em arXiv preprint arXiv:1412.6980}, 2014.

\bibitem{kramer2013k}
O.~Kramer and O.~Kramer, ``K-nearest neighbors,'' {\em Dimensionality reduction with unsupervised nearest neighbors}, pp.~13--23, 2013.

\end{thebibliography}
\end{document}